\documentclass[a4paper,11pt]{article}
\usepackage{pos}
\usepackage{bm}
\usepackage{float}
\usepackage[utf8]{inputenc}
\newcommand{\MM}{\mathcal{M}}
\newcommand{\cQ}{\mathcal Q}
\newcommand{\cT}{\mathcal{T}}
\newcommand{\cC}{\mathcal{C}}
\def\d {{\rm d}}

\newcommand{\be}{\begin{equation}}
\newcommand{\ee}{\end{equation}}
\newcommand{\beq}{\begin{equation}}  
\newcommand{\eeq}{\end{equation}}

\def\d {{\rm d}}
\def\del          {\partial}

\def\ii           {{\rm i}}

\def\Im           {{\rm Im\hskip0.1em}}

\def\calb         {{\cal B}}
\def\calc         {{\cal C}}

\def\calg         {{\cal G}}

\def\calq         {{\cal Q}}

\title{Large Field Distances from EFT strings}

\author[a]{Stefano Lanza}
\author[b]{Fernando Marchesano}
\author*[c]{Luca Martucci}
\author[b,d,e]{Irene Valenzuela}

\affiliation[a]{Institute for Theoretical Physics,
Utrecht University, 
Princetonplein 5, 3584 CE Utrecht, The Netherlands}

\affiliation[b]{Instituto de F\'{\i}sica Te\'orica UAM-CSIC, c/ Nicol\'as Cabrera 13-15, 28049 Madrid, Spain}

\affiliation[c]{Dipartimento di Fisica e Astronomia “Galileo Galilei”, Universit\`a degli Studi di Padova\\
\& I.N.F.N. Sezione di Padova, Via F. Marzolo 8, 35131 Padova, Italy}

\affiliation[d]{Departamento de F\'{\i}sica Te\'orica, Universidad Aut\'onoma de Madrid, Cantoblanco, 28049 Madrid, Spain}

\affiliation[e]{CERN, Theoretical Physics Department, 1211 Meyrin, Switzerland}

\emailAdd{s.lanza@uu.nl}
\emailAdd{fernando.marchesano@csic.es}
\emailAdd{luca.martucci@pd.infn.it}
\emailAdd{irene.valenzuela@uam.es}

\abstract{In any consistent effective field theory of quantum gravity limits of infinite field distance are expected to lead to the EFT breakdown due to the appearance of an infinite tower of light states, as predicted by the Distance Conjecture. We review the Distant Axionic String Conjecture, which proposes that any 4d EFT infinite-field-distance limit  can be realized as an RG flow  of a fundamental axionic string. The RG flow can be understood in terms of the 4d backreaction of such a string, and implies that it becomes tensionless towards the said limit. This property is understood as a shielding mechanism towards realizing an exact axionic symmetry, and it implies the breakdown of the EFT in a way that reproduces the Distance Conjecture. Motivated by string theory data we further propose the Integral Scaling Conjecture, which provides a specific relation between the string tension and the EFT maximal cut-off set by the infinite tower of states.}

\FullConference{%
  Corfu Summer Institute 2021 "School and Workshops on Elementary Particle Physics and Gravity"\\
  29 August - 9 October 2021\\
  Corfu, Greece
}

\begin{document}
\maketitle

\section{Introduction}

In physical models it is  typically hard to extract exact physical information. A standard approach is to exploit the presence of some symmetries or to work perturbatively with respect to some small parameter $g\ll 1$. However, these strategies can be applied only in a peculiar and restrictive sense in quantum gravity (QG). Indeed, the widely accepted  no global symmetry conjecture \cite{Banks:1988yz,Banks:2010zn} claims that exact global symmetries are not admitted in QG. Furthermore, string theory clearly points to the absence of freely tunable constants  which could be used as expansion parameter $g$. This implies that any physical coupling of a QG model should be rather considered as a function $g(\phi)$ of some field $\phi$, and is then  determined by the dynamics of $\phi$. Hence, a non-accidental global symmetry can be preserved only in some approximate sense, and this requires a specification of a perturbative regime, which can in turn be realized only in some field space regions. Understanding how the correlation between these aspects can generically affect an effective field theory (EFT) consistent with QG is clearly of fundamental importance, from both the purely theoretical and the phenomenological points of view.   

\begin{figure}[tb]
	\centering
	\includegraphics[width=9cm]{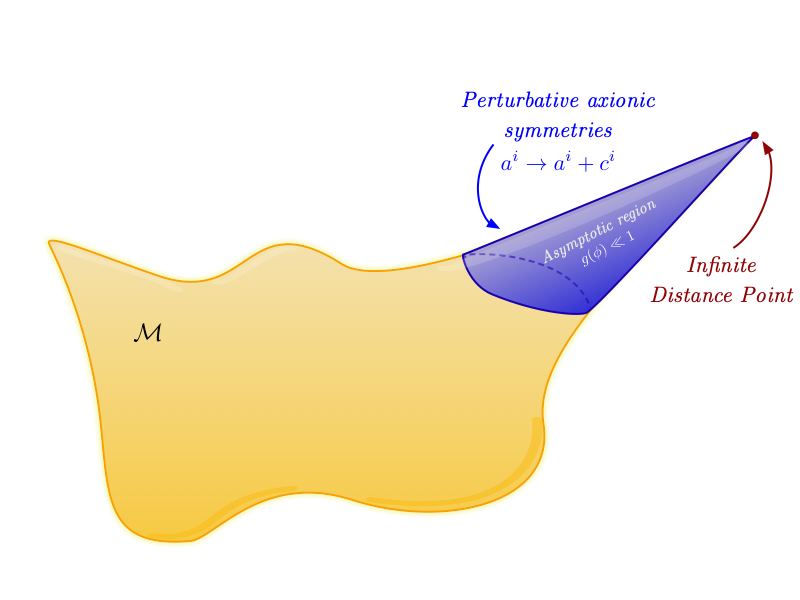}
	\caption{Within the asymptotic region of the field space $\mathcal{M}$, depicted in blue, the continuous shift symmetries $a^i \to a^i + c^i$ are approximately  preserved. \label{f:AsymptoticStringsRegimes}}   
	\end{figure}

In this note we review some recent progress in this direction \cite{Lanza:2020qmt,Lanza:2021udy}, which focuses on the axionic sectors of four-dimensional EFTs. In particular, we are interested in {\em fundamental} axions $a^i\simeq a^i+1$, of the kind coming from the zero-modes of some higher-dimensional gauge field in a string compactification. These are more easily characterized in terms of the corresponding axionic strings, around which the axions undergo an integral shift set by the string charges $e^i$: $a^i\rightarrow a^i+e^i$. These strings are {\em fundamental} in the sense that they cannot be resolved into a smooth solitonic object within a four-dimensional EFT -- see also \cite{Reece:2018zvv}. Hence these sectors more directly depend on the UV completion  of the EFT, and are then particularly sensitive to QG effects. Furthermore, by the above observations, the axionic shift symmetries $a^i\rightarrow a^i+c^i$ can be at best {\em perturbatively} preserved, as it is always broken by exponentially suppressed non-perturbatively  corrections  $\sim e^{-A/g^2(\phi)}$. This implies that these fundamental axions and strings  really characterize some {\em asymptotic} field-space regions where $g(\phi)\ll 1$, around some infinite distance field-space boundary at which the axionic shift symmetries become formally exact -- see Figure~\ref{f:AsymptoticStringsRegimes} for a pictorial representation.

In these asymptotic regions, the fundamental axionic strings can be regarded as fundamental  electrically charged objects, coupled to the two-form potentials $\calb_{2,i}$ dual to the axions $a^i$ through $e^i\int {\cal B}_{2,i}$. Hence, by invoking the completeness hypothesis \cite{Polchinski:2003bq} one can argue that, in fact, fundamental strings {\em must} be included into the EFT. All this suggests that fundamental axionic strings could play a distinguished role as probes of these asymptotic field-space regions, as well as of the physical properties of the associated perturbative EFT regimes. In this note we will explain how this expectation can be concretely confirmed, by focusing on the more controlled (but still quite flexible) EFTs preserving minimal supersymmetry at their  UV cut-off scale. In particular, we will see how a special subclass of axionic strings, the {\em EFT strings}, provide a natural way to characterize the possible perturbative regimes of a given EFT admitting a UV QG completion. It is widely believed that such perturbative regimes correspond to limits of infinite distance in field space, and that QG enforces a protection mechanism against reaching them physically. The general proposal describing such a protection mechanism is the Swampland Distance Conjecture (SDC) \cite{Ooguri:2006in} which predicts an exponential fall-off for the mass scale $m_*$ of an infinite tower of states as we reach any point at infinite distance in field space. The importance of this statement is that $m_*$ represents the maximal cut-off of the EFT: a new, unexpected scale from the pure EFT viewpoint. This sort of statement is also predicted by the magnetic Weak Gravity Conjecture (WGC) for particles \cite{ArkaniHamed:2006dz}, in those limits where a gauge coupling $g$ vanishes. The relationship between these two statements is particulary significant in certain contexts, like in 4d ${\cal N}=2$ vector multiplet moduli spaces, as illustrated in \cite{Grimm:2018ohb,Gendler:2020dfp}.

As we will see, one can  provide a direct link between the WGC and the SDC, allowing for a bottom-up derivation of the exponential fall-off for $m_*$. The link is built for the case of 4d EFTs, through two proposals dubbed Distant Axionic String Conjecture (DASC) and Integral Scaling Conjecture (ISC) in \cite{Lanza:2020qmt,Lanza:2021udy}. The DASC  provides a physically distinguished way to reach infinite distance points in terms of EFT string  flows. Combined with the WGC for strings, it also implies that the tension of the EFT string generating the flow provides a maximal cut-off with an exponential drop-off, as predicted by the SDC. This maximal cut-off is not necessarily $m_*$, but the relation between both turns out to be simply described by an integer in a plethora of string theory constructions, which prompts us to formulate this as a general principle captured by the ISC. Together, these two conjectures relate the exponential drop-off for $m_*$ with the extremality factor of the corresponding EFT string.

\section{String flows and EFT strings}
\label{sec:2}

\noindent In the previous section we have argued that in quantum gravity  fundamental axionic strings are natural candidates to probe asymptotic field-space regions  of an  EFT. Since we are interested in four-dimensional theories, however, one faces a  complication. In four dimensions, strings are  codimension-two objects and cannot generically be considered as probes. For instance, a straight string typically destroys any asymptotic vacuum. However, as we are going to explain, these strong backreaction effects are not really an issue for our purposes, and will turn out to be the key for identifying the class of strings which are more directly sensitive to the UV completion of the EFT.

Henceforth  we will focus on four-dimensional EFTs preserving minimal ${\cal N}=1$ supersymmetry at the UV cut-off scale $\Lambda$.  Of course, supersymmetry can be spontaneously broken at lower energies, but our considerations will regard the EFT structure at energy scales of order $\Lambda$. The general  ${\cal N}=1$ supergravity including BPS strings (and membranes) has been discussed in detail in \cite{Lanza:2019xxg}. Here we recall just a few  ingredients -- see \cite{Lanza:2019xxg} for more details.   

The axions $a^i$ are completed into the bottom components $t^i=a^i+\ii s^i$ of chiral superfields, where $s^i$ are the {\em saxions}. We assume that the EFT K\"ahler potential $K$ depends on the chiral fields $t^i$ only through their saxionic components $s^i=\Im t^i$, as it is appropriate for describing asymptotic field-space regions in which the axionic shift symmetries are perturbatively preserved.  A  BPS string carrying a ${\cal B}_{2,i}$ charge vector ${\bf e}=\{e^i\}$ has a field-dependent tension
\be\label{thetension}
\cT_{\bf e}\equiv M^2_{\rm P}\,e^i\ell_i
\ee
where $\ell_i(s)\equiv -\frac12 \frac{\partial K}{\partial s^i}$ are the {\em dual saxions}, which provide an alternative parametrization of the saxionic directions.

	\begin{figure}[th]
		\centering
		\includegraphics[width=9cm]{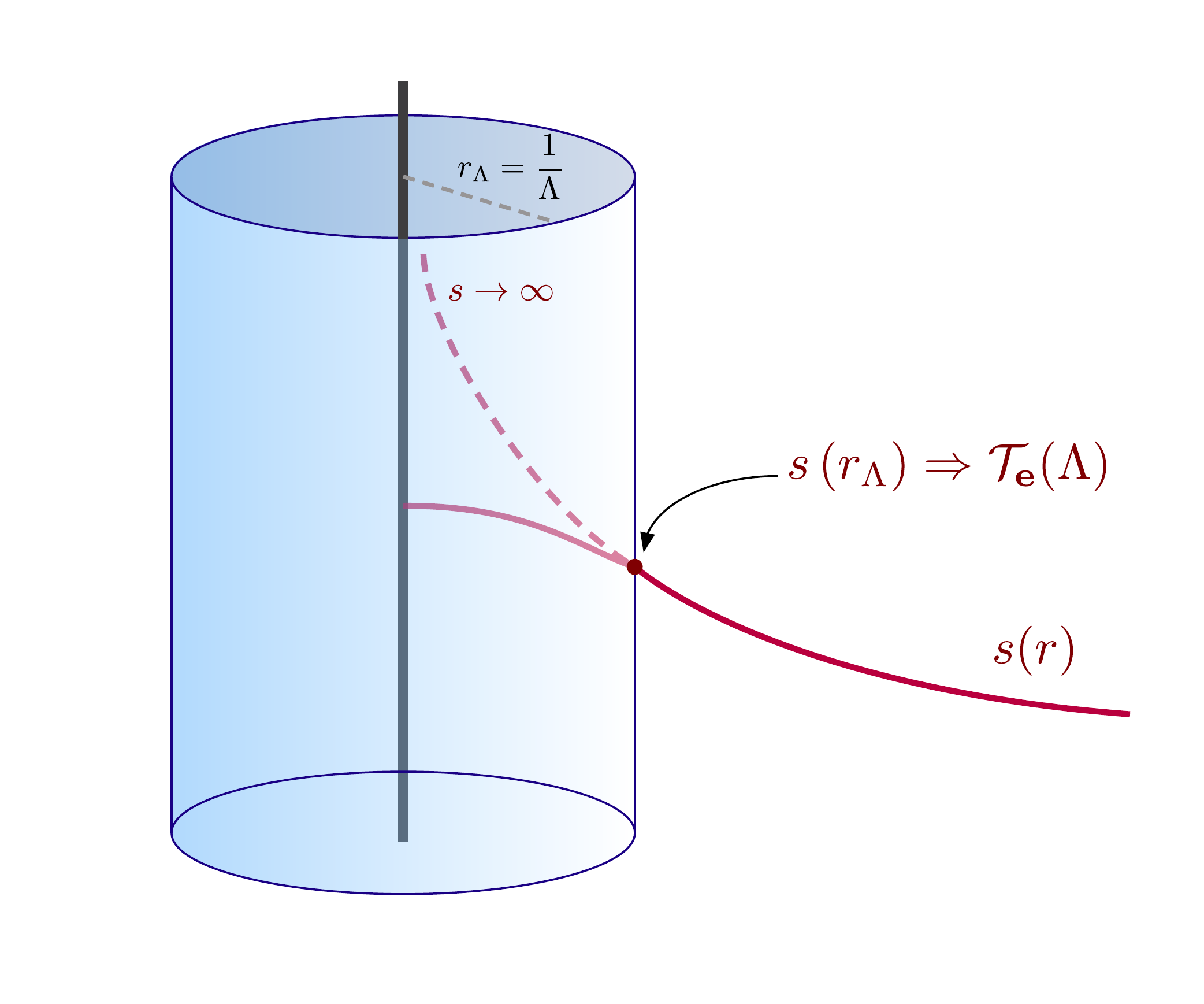}
		\caption{Given an EFT with cutoff $\Lambda$, one cannot resolve spacetime regions at distances smaller than $ r_\Lambda \equiv \Lambda^{-1}$. The EFT string tension is defined as the string tension $\cT_{\bf e}(\Lambda)$ measured at the cutoff scale $\Lambda$, where the saxionic fields have vacuum expectation value $s(r_{\Lambda})$. \label{f:RG_Strings}}   
	\end{figure}

As an example of the above-mentioned backreaction issue,  we will presently see that the presence of a string forces the saxions to flow around it, and to actually degenerate at the string location. This seems to make problematic  an appropriate interpretation of the tension \eqref{thetension}. However, it is important to keep in mind that the bulk-plus-string EFT  is defined at the cut-off scale $\Lambda$, which sets the coarse-grain length-scale $r_{\Lambda}\equiv \Lambda^{-1}$. Hence, the string tension \eqref{thetension} should be considered as cutoff-dependent and  should be roughly set  by the value of the bulk scalars at a distance $r_{\Lambda}$ from the string location:  $\cT_{\bf e}(\Lambda)\equiv   M^2_{\rm P}\,e^i\ell_i(r_{\Lambda})$. In particular, this implies that  from the viewpoint of a 4d EFT $\cT_{\bf e}(\Lambda)$ changes as we change $\Lambda$ because of the string backreaction, which can be interpreted as an RG-flow effect \cite{Lanza:2020qmt,Lanza:2021udy}, along the lines of \cite{Goldberger:2001tn,Michel:2014lva,Polchinski:2015bea} -- see Figure~\ref{f:RG_Strings}.

	\begin{figure}[t]
		\centering
		\includegraphics[width=10cm]{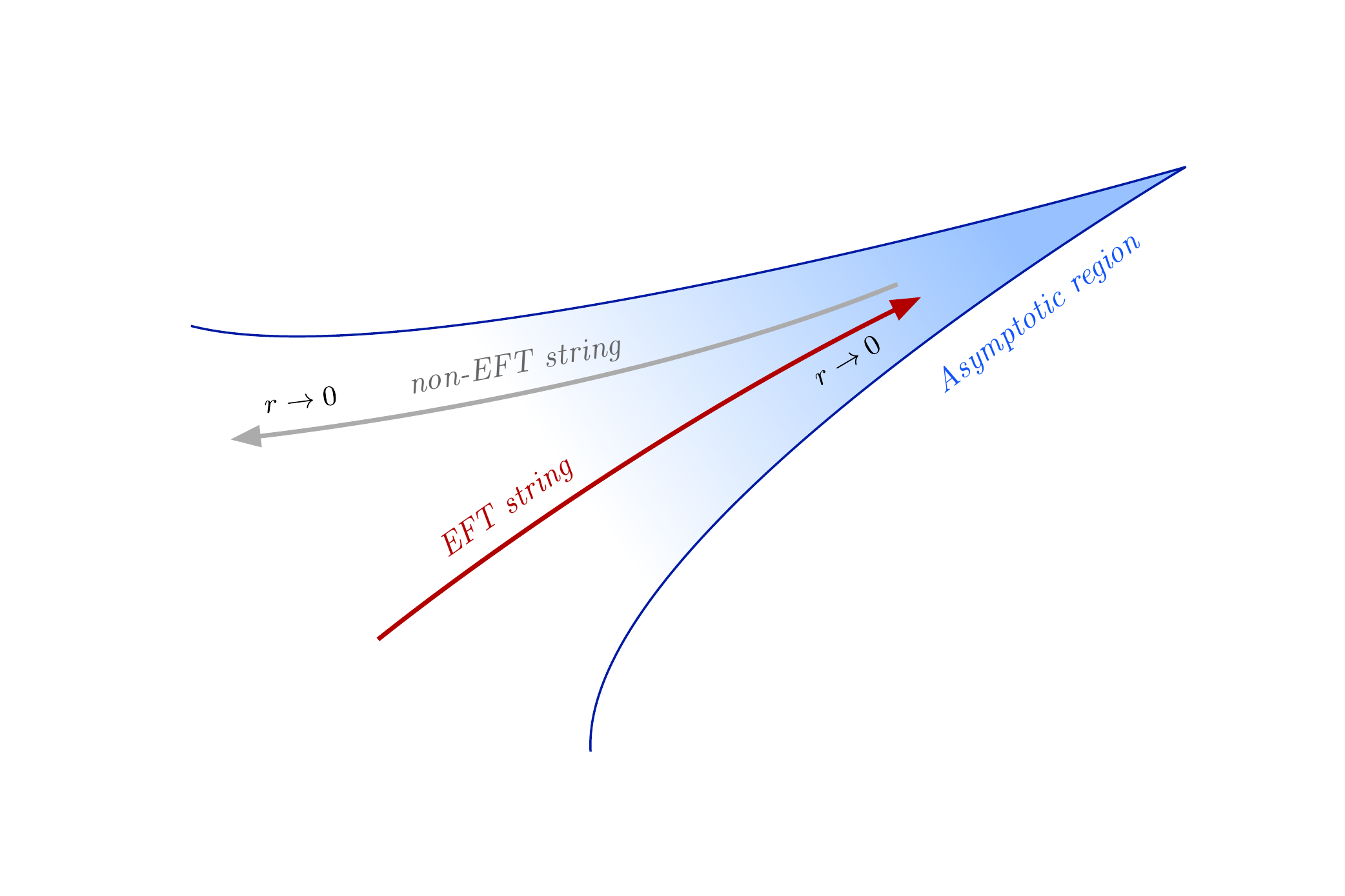}
		\caption{The flow of EFT strings drives the scalar fields towards asymptotic field-space regions where the EFT admits a perturbative regime and corresponding axionic shift symmetries; vice versa, the flow of a non-EFT string drives the fields towards regions where this perturbative regime breaks down. \label{f:EFT_non-EFT_Strings}}   
	\end{figure}

So far we required the relevant bulk sector to be in an appropriate asymptotic field-space region, in which we can consider the axionic symmetries as perturbatively preserved. However, it is sufficient to assume this only as we approach the string, up to the cut-off distance $r_{\Lambda}$. We are then  led to focus on the special subclass of  BPS strings whose backreaction automatically drives the surrounding scalars towards the appropriate asymptotic region.  These strings naturally allow for a controlled weakly-coupled EFT description and for this reason  in \cite{Lanza:2020qmt,Lanza:2021udy} they have been dubbed {\em EFT strings}. BPS strings not sharing this property will be referred to as {\em non-EFT} strings -- see Figure~\ref{f:EFT_non-EFT_Strings}.

This qualitative distinction between EFT and non-EFT strings can be made more precise by further exploiting supersymmetry. Indeed, we are interested in the string backreaction as we approach the  cut-off distance $r_{\Lambda}$. At such length scales  the string is assumed to (approximately) preserve  one-half of the bulk supersymmetry and, independently of its macroscopic shape, its core locus can be well approximated by a straight line. Furthermore, we can neglect a possible low-energy potential and treat the scalars as massless. Hence, the string backreaction can be described by a $\frac12$-BPS flow of the kind discussed for instance in \cite{cstring,Dabholkar:1990yf}. In particular, the chiral fields $t^i(z)$ must depend holomorphically on the transverse complex coordinate $z$  \cite{cstring}. Close to the BPS string, the leading contribution to its backreaction is completely fixed by its charges $e^i$:
\be\label{tsol}
t^i\simeq t^i_{0}+\frac{e^i}{2\pi\ii}\log\frac{z}{z_0}
\ee
where we have assumed that the string is located at $z=0$, and $t^i_{0}$ represent arbitrary initial values at $z_0$. One can also explicitly check how, in full generality, such a backreaction induces a consistent RG-flow of the string tension $\cT_{\bf e}(\Lambda)$ as defined above \cite{Lanza:2021udy}.

The string flow \eqref{tsol} realizes the correct axion-shift $a^i\rightarrow a^i+e^i$ around it. But most importantly for us, it shows how the  saxions  flow along straight lines generated by the charge vector ${\bf e}=\{e^i\}$, in the saxionic plane:
\be\label{sflow}
{\bm s}(\sigma)={\bm s}_0+{\bf e}\, \sigma\quad,\quad~~~\text{with}\quad\sigma\equiv \frac1{2\pi}\log\frac{r_0}{r}\,, 
\ee
where ${\bm s}=\{s^i\}$. 

We can now use this explicit description to provide a more quantitative characterization of the EFT strings. In order to do that, we need to specify how the perturbative regime associated with the axionic shift symmetries $a^i\rightarrow a^i+c^i$ is defined. We  make the reasonable assumption that the leading non-perturbative effects breaking the axionic symmetries are generated by BPS instantons, which contribute to the EFT by terms proportional to $e^{2\pi m_it^i}$, where $m_i$ denote the BPS instanton charges. The strength of these non-perturbative corrections is then set by 
\be 
\left|e^{2\pi m_it^i}\right|=e^{-2\pi\langle {\bf m},{\bm s}\rangle}
\ee 
where we have introduced the natural pairing $\langle {\bf m},{\bm s}\rangle\equiv m_i s^i$ between instanton charges and saxions. The perturbative regime is hence  associated to a given set $\calc_{\rm I}$ of  BPS instanton charges such that $e^{-2\pi\langle {\bf m},{\bm s}\rangle}\ll 1$ for any ${\bf m}\in \calc_{\rm I}$. The perturbative regime then corresponds to the requirement  that the saxions lie deep inside the {\em saxionic cone} $\Delta\equiv \{{\bm s}\ |\ \langle {\bf m},{\bm s}\rangle > 0\,,\ \forall {\bf m}\in \calc_{\rm I}\}$.   

We then arrive at the following identification of the possible EFT string charges. Let us denote by $\calc_{\rm S}$ the set of possible BPS string charges. Then the subset of possible EFT string charges is given by
\be\label{EFTcharge}
\calc^{\text{\tiny EFT}}_{\rm S}\equiv \{{\bf e}\in \calc_{\rm S}\ |\ \langle {\bf m},{\bf e}\rangle \geq 0\,,\ \forall {\bf m}\in \calc_{\rm I}\}\subset \calc_{\rm S}\,.
\ee
Indeed, along the saxionic flow equation \eqref{sflow} we have $e^{-2\pi\langle {\bf m},{\bm s}\rangle}=e^{-2\pi\langle {\bf m},{\bm s}_0\rangle}e^{-2\pi\langle {\bf m},{\bm e}\rangle\sigma}$. However, this is  exponentially suppressed for any ${\bf m}\in \calc_{\rm I}$ as $\sigma\rightarrow \infty$, i.e.\ as we approach the string,  if and only if ${\bf e}\in \calc^{\text{\tiny EFT}}_{\rm S}$.\footnote{Note that the definition \eqref{EFTcharge} shares some similarities with  the definition of `supergravity strings'  in $d\geq 5$ dimensions proposed in \cite{Kim:2019vuc,Katz:2020ewz} -- see \cite{Lanza:2021udy} for more comments on this analogy.}

\begin{figure}[tb]
		\centering
		\includegraphics[width=12cm]{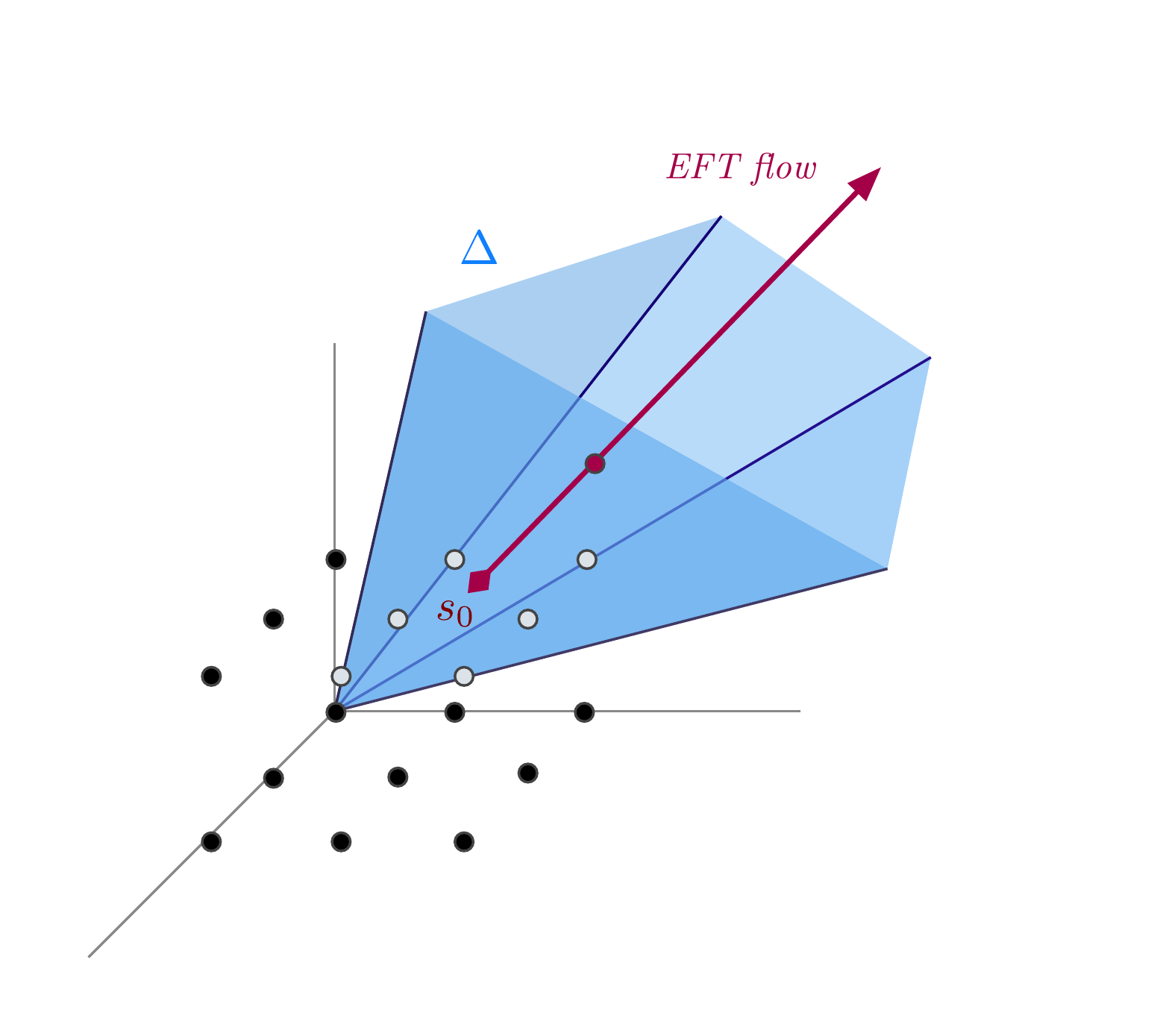}
		\caption{The saxionic cone $\Delta$ can be regarded as generated by EFT string charges (in white and red). Any EFT string charge (like the red one in the picture) generates a flow that, starting from ${\bf s}_0 \in \Delta$, is fully contained within $\Delta$. \label{f:Cone_EFT_Flow}}   
	\end{figure}

Note that $\calc^{\text{\tiny EFT}}_{\rm S}$ can be regarded as a discretization of the closure $\overline\Delta$ of the saxionic cone. Hence, from a more geometrical viewpoint, a  flow \eqref{sflow}   generated by an EFT string charge ${\bf e}\in \calc^{\text{\tiny EFT}}_{\rm S}$ can be identified with a straight line starting from some point ${\bm s}_0\in \Delta$ and  remaining inside $\Delta$ for any  $\sigma\in[0,\infty)$ -- see Figure~\ref{f:Cone_EFT_Flow}.  Since  the  axionic shift symmetries become asymptotically exact as  $\sigma\rightarrow\infty$ along a flow \eqref{sflow} generated by an EFT string charge ${\bf e}\in \calc^{\text{\tiny EFT}}_{\rm S}$, this flow must represent a limit to infinite distance in field space.

\section{EFT strings, WGC and SDC}

\noindent In the previous section we argued that, in a given perturbative regime, the flows that are generated by EFT strings drive the saxionic fields towards infinite distance limits. 
The Distant Axionic String Conjecture (DASC) proposes that also the reverse holds, stating that \emph{all} points  on the infinite distance boundary of field space can be regarded as an endpoint of an EFT string flow:

\vspace{1em}
\fbox{\parbox{0.85\linewidth}{\centerline{\bf Distant Axionic String Conjecture}
\bigskip
{\em\noindent Every infinite field distance point of a four-dimensional EFT consistent with quantum gravity can be  realized as the  endpoint  of an EFT string flow.}}}
\vspace{1em}

\noindent To properly understand this proposal, one should make use of a triple correspondence between EFT string flows, backreacted solutions and paths in field space $\MM$. Using this correspondence, an EFT string of charge vector {\bf e} generates a path in $\MM$ that has the form \eqref{sflow} when $\sigma=\frac{1}{2\pi}\log\frac{r_0}{r}$ is varied. As pointed out in the previous section, by identifying $r$ with $r_{\Lambda}=\Lambda^{-1}$, this flow can be regarded as a world-sheet RG-flow associated with the variation of  the EFT cut-off $\Lambda$. From this viewpoint, infinite distance loci are reached in the limit $\Lambda \to \infty$; however, it is worth stressing that this limit is only formal, for the 4d effective description stops being valid at some intermediate scale.

This proposal is quite natural in a minimally supersymmetric setting, and  basically amounts to saying that all possible infinite distance regimes are associated with fundamental axionic symmetries. By the Completeness Hypothesis we would then expect that fundamental string operators with the corresponding magnetic charges appear in that regime of our EFT, and everything else follows from this. This conjecture can also be easily extended to non-supersymmetric setups whenever supersymmetry is spontaneously broken and restored at some scale $\Lambda_{\rm SUSY}$ below the EFT cut-off. Instead, if supersymmetry is restored at some energy scale higher than the EFT cutoff,  the realization of the DASC is non-trivial. In particular, the DASC would require that every scalar field whose trajectory parametrizes the infinite field distance path is paired with an axion-like field that is magnetically coupled to the string realizing said trajectory. Therefore, the perspective delivered by the DASC can be regarded as the opposite of Conjecture 4 in \cite{Ooguri:2006in}, according to which every axion must be accompanied by a radial mode, namely a saxion, in order to ensure the vanishing of first homotopy group of the moduli space.

The Swampland Distance Conjecture (SDC) \cite{Ooguri:2006in} predicts that along any geodesic infinite distance limit an infinite tower of new  light states appears, whose lightest mass $m_*$ decreases exponentially as $e^{-\alpha\Delta\phi}$ with the field distance $\Delta\phi$, for some constant $\alpha$. 
From the viewpoint of the EFT, there is a priori no way to identify neither such {\it tower scale} $m_*$ nor the corresponding constant $\alpha$ without knowing the EFT UV completion. However, the above proposal implies a physical criterium to classifying the infinite distance limits as EFT-string-generated flows \eqref{sflow}. The flow also dictates the renormalization of the EFT string tensions, which provide a candidate for the maximal UV cut-off.  Indeed, it is natural to assume that the EFT strings themselves provide a tower of states like the one predicted by the SDC, as its presence is a natural shielding mechanism against fundamental axionic symmetries. The advantage of this scheme is that such a maximal EFT breaking scale can be computed with EFT data.

Under this assumption, one can provide a bottom-up derivation of the SDC in 4d EFTs via the DASC and the Weak Gravity Conjecture (WGC) for strings. The crucial point is that an EFT string tension goes to zero monotonically along its RG flow, so the string becomes lighter and lighter as we approach the infinite distance limit in moduli space. When the string tension gets below $\Lambda^2$, the EFT necessarily breaks down. One can then show that the exponential behavior of the EFT-breaking scale becomes just a consequence of the WGC for EFT strings.

To see the relation between the WGC and the SDC, we have first to provide a way to formulate the WGC for strings in four dimensions, which is not obvious for the very same backreaction issues  discussed at the beginning of section \ref{sec:2}. However, a way out is again to focus on quantities appearing directly in the EFT, considering them as defined at the EFT cut-off scale $\Lambda$. One can then define a `physical' string charge as follows
 \be\label{calgdef}
\calq_{\bf e}=M_{\rm P}\sqrt{ \calg_{i j}e^i e^ j}\, ,
\ee
where $\calg_{i j}\equiv \frac12 \frac{\del^2 K}{\del s^i\del s^j}$ is the (s)axion kinetic matrix, which is equals to the inverse of the $\calb_{2,i}$ kinetic matrix and then measures the $\calb_{2,i}$ gauge couplings. Now a WGC bound for strings can be written as follows: 
 \be\label{WGC}
M_{\rm P}\calq_{\bf e}\geq \gamma \cT_{\bf e}\, ,
\ee
where $\gamma$ is a constant. As discussed in \cite{Lanza:2020qmt}, one can explicitly check that \eqref{WGC} is indeed satisfied by EFT strings in several string theory models.

Consider now the BPS saxionic flow \eqref{sflow}. Combining it with \eqref{thetension} and \eqref{calgdef}, one can easily get 
\be\label{monoT}
\frac{\d\cT_{\bf e}(\sigma)}{\d\sigma}=-\cQ^2_{\bf e}<0\, ,
\ee 
where $\cT_{\bf e}$ and $\cQ^2_{\bf e}$ are evaluated at $s^i(\sigma)$. We then see that supersymmetry alone is sufficient to conclude that the tension of a string must decrease alone its own flow. Furthermore, \eqref{monoT}  allows us to write the overall field-space distance traveled along a string flow in terms of the corresponding tension variation. Indeed, by combining \eqref{sflow}, \eqref{calgdef} and \eqref{monoT} one obtains 
\be\label{dT}
\mathrm{d}_\sigma =\frac{1}{M_{\rm P}}\int_0^{\sigma} \cQ_{\bf e}\d\sigma= \frac{1}{M_{\rm P}}\int^{\cT^0_{\bf e}}_{\cT_{\bf e}(\sigma)} \frac1{\cQ_{\bf e}}\d\cT_{\bf e}\, .
\ee
We can now combine the WGC bound \eqref{WGC} with \eqref{dT} to get the bound $\mathrm{d}_\sigma\leq  \frac1{\gamma} \log\frac{\cT^0_{\bf e}}{\cT_{\bf e}(\sigma)}$.
Thus, the maximal EFT-breaking scale that is consistent with the existence of the string falls off as
\beq
\label{Lmax}
\Lambda_{\rm max}^2=\cT_{\bf e}(\sigma)\leq  \cT_{\bf e}^0 \exp \left(-\gamma\  \mathrm{d}_\sigma \right)\, .
\eeq
In words, the WGC implies the exponential drop-off of the maximal cut-off predicted by the SDC along every field space path corresponding to an EFT string flow. Then, using  the DASC, one can argue that this finding is universal.

As discussed in \cite[appendix C]{Lanza:2021udy}, it can be further shown that the EFT string flows are asymptotically geodesic. Therefore the above derivation also provides a relation between the exponential rate of the SDC and the EFT string extremality factor $\gamma$. In fact this relation is an upper bound, as in general the {\em microscopic} tower scale $m_*$ could be below $\sqrt{\cT_{\bf e}}$. A more precise relation can be derived by invoking yet another conjecture, that we now turn to discuss.

\section{Microscopic scales and integral scaling}

We have illustrated how the DASC can be employed to realize infinite distance limits, and how these can be classified by exploiting the physical properties of EFT strings. Indeed, the charge of the EFT strings determines the specific form of the infinite distance paths \eqref{sflow}. Moreover, the EFT string flows dictate the renormalization of the string tensions which provide a maximal EFT cut-off scale. While  the EFT strings themselves may provide the leading infinite tower of states that signal the EFT breakdown, in general there are additional towers of microscopic states becoming lighter at a rate faster than the EFT string. One may then wonder how the EFT string tension $\cT$ and the leading microscopic tower scale $m_*$ are related. By direct inspection of a large number of string/M-theory models, one observes a remarkable correlation between the asymptotic scaling of $\cT$ and $m_*$, which has been proposed as a general property in \cite[Conjecture 2]{Lanza:2020qmt}. Here we propose a slightly stronger version thereof:

\vspace{1em}
\fbox{\parbox{0.85\linewidth}{\centerline{\bf Integral Scaling Conjecture}
\bigskip
{\em\noindent  Along the asymptotic flow associated with an EFT string, its tension $\cT$ goes to zero. The microscopic tower mass $m_*$ then scales like
\be\label{m*}
m_*^2\simeq M^2_{\rm P}A\left(\frac{\cT}{M^{2}_{\rm P}}\right)^{{\it w}}\quad,\quad {\it w}\in \{1,2,3\}
\ee 
with $A$ a coefficient not  depending  on the flowing fields. }}}
\vspace{1em}

\noindent We refer to the integer ${\it w}$ as the {\it scaling weight}  of the EFT string. The only difference with respect to \cite[Conjecture 2]{Lanza:2020qmt} is that therein the scaling weight could be any positive integer. For this reason, we can  refer to the proposal of \cite{Lanza:2020qmt} as the weak form of the  Integral Scaling Conjecture.
Here we are promoting to a general principle the   observation that, in all   string theory examples considered so far, one finds only ${\it w}=1,2,3$ as possible values for the scaling weight. One also finds $A\simeq \mathcal{O}(1)$ for generic values of the non-flowing fields.

The scaling weight ${\it w}$ is related to the exponential rate $\alpha$ that appears in the formulation of the SDC. In fact, due to \eqref{m*} the exponential behavior in \eqref{Lmax} implies an exponential drop-off for $m_*$:
\beq
m_*\leq m_*^0\exp(-\alpha\,{\rm d}_\sigma )\ , \quad \alpha=\frac{{\it w}\gamma}2\, ,
\eeq
with $\gamma$ the constant appearing in \eqref{WGC}.

Furthermore in all considered string theory  examples we have observed the following intriguing convexity  of the scaling weight, regarded as a function ${\it w}_{{\bf e}}$ of the EFT string charge vector ${\bf e}$:  
\be\label{covexindex}
{\it w}_{{\bf e}_1+{\bf e}_2}\leq {\it w}_{{\bf e}_1}+{\it w}_{{\bf e}_2} \quad~~~~~~\forall {\bf e}_1,{\bf e}_2\in \cC_{\rm S}^{\text{\tiny EFT}}\,.
\ee

\section{Final remarks}

Both the Distant Axionic String Conjecture and the Integral Scaling Conjecture are supported by a plethora of infinite distance limits of 4d string theory constructions, as shown in \cite[section 6]{Lanza:2020qmt}. In particular, they were tested in those EFTs obtained from:
\begin{itemize}

\item[-] Heterotic strings compactified on Calabi--Yau three--folds;
\item[-] Type I string theory compactified on Calabi--Yau three--folds;
\item[-] Type IIB/F-theory compactifications to 4d;
\item[-] Type IIA strings compactified on Calabi--Yau orientifolds with O6-planes;
\item[-] M-theory compactified on $G_2$ manifolds. 

\end{itemize}
All these constructions have in common that just below the compactification scale one recovers a 4d ${\cal N}=1$ theory, which may then be broken down to ${\cal N} =0$ at lower scales. As mentioned before, a highly non-trivial test of these conjectures would be to consider them in ${\cal N} =0$ compactifications with light scalar fields and where supersymmetry is not recovered at the compactification scale. 

In any event, these encouraging results motivate exploring the consequences of the DASC and ISC with respect to the web of swampland conjectures. While a direct relationship between the WGC for strings and the SDC is perhaps the most remarkable one, there are other consequences as well. One of them is the classification of different infinite distance points in terms of EFT string flows, and more precisely in terms of an integer $p$ dubbed degeneracy index of the flow. In short, when translated into dual variables $\ell_i$ the saxionic cone is mapped to a dual saxionic cone ${\cal P}$. Any EFT string flow must end on one of the boundaries ${\cal P}$, and $p$ specifies the codimension of such a boundary. While the physical significance of this degeneracy index is not fully understood yet, it plays an important role in the whole structure of EFT string charges unveiled in \cite{Lanza:2020qmt}. It also seems to have a direct link to the curvature of the moduli space around infinite distance points, which is one of the classical questions in the Swampland Program. Indeed, on the one hand, one can argue that the endpoints of non-degenerate flows (those with $p=1$) correspond to regions of negative scalar curvature, as proposed in \cite[Conjecture 3]{Ooguri:2006in}. On the other hand, degenerate flows (those with $p>1$) contain counterexamples to \cite[Conjecture 3]{Ooguri:2006in}. Finally, a direct implication of the DASC is that around each infinite distance point  there is at least one holomorphic sectional curvature that is negative, which is a much weaker property than a non-positive scalar curvature. 

It is also worth mentioning that the DASC and ISC fit nicely with the Emergent String Conjecture (ESC) \cite{Lee:2019wij}, which distinguishes two classes of infinite distance limits: decompactification limits or emergent critical string limits. Applied to our 4d setup, one would identify asymptotic limits with a scaling weight $w > 1$ with decompactification limits, while those with scaling weight $w = 1$ should contain the limits in which the EFT string is an emergent critical string. In fact, the ESC provides a rationale for the range $1 \leq w \leq 3$ proposed in the ISC. Indeed, a value $w<1$ would correspond to a limit in which we would find a 4d critical string, in direct tension with common wisdom. Additionally, one can show that if $w > 3$ and $m_*$ corresponds to a KK scale, then the string tension lies above the species scale. This would be quite unnatural from the perspective of the Emergence Proposal \cite{Grimm:2018ohb,Heidenreich:2018kpg}, since it would mean that the axion that couples magnetically to the EFT string is still fundamental above the species scale, and so infinite distance points should also exist at this quantum gravity scale. Furthermore, let us notice that determining $w$ within the EFT would allow us to estimate the EFT cut-off only by employing EFT data!\footnote{The EFT string perspective  also fits nicely  with the  holographic moduli space reconstruction of \cite{Grimm:2020cda,Grimm:2021ikg}, where the knowledge of some boundary data permits to reconstruct the moduli space in the near-boundary region. In fact, the global string solutions of \cite{Marchesano:2022avb} suggest that EFT string data could encode information about the whole moduli space.} 

Finally, EFT strings can also be understood as the cobordism defect required to break the global symmetry associated to having non-trivial 1-cycles in field space (i.e. the dual 2-form global symmetry of the axions). Hence, a generalization of the DASC to higher dimensions would likely involve the identification of infinite distance limits with the RG flows induced by the corresponding cobordism defects, which should anyway exist according to the Cobordism Conjecture \cite{McNamara:2019rup}. This idea has been further explored in \cite{ Buratti:2021fiv}  where it was indeed proposed that any infinite distance limit can be realized as the running into a cobordism wall of nothing.

Our results highlight the relevance of extended objects to understand the principles underlying the SDC, even if the leading tower of states becoming light does not necessarily come from them. A key principle to unveil all these features is the map between large field distances in field space and large field distances induced by low-codimension objects in an EFT, in a spirit analogous to   \cite{Draper:2019utz}. This perspective might hopefully lead to  a bottom-up rationale for the SDC, independently of its string theory realization.





\providecommand{\href}[2]{#2}\begingroup\raggedright\endgroup

\end{document}